\documentclass[aps,prd,twocolumn,groupedaddress]{revtex4}

\usepackage{graphicx}
\usepackage{dcolumn}

\begin{document}

\preprint{UTAP-560}

\title{Dark matter annihilation from intermediate-mass black holes: \\Contribution to the extragalactic gamma-ray background}


\author{Shunsaku Horiuchi}
\email[]{horiuchi@utap.phys.s.u-tokyo.ac.jp}
\affiliation{Department of Physics, School of Science, The University of Tokyo, Tokyo 113-0033, Japan}

\author{Shin'ichiro Ando}
\email[]{ando@tapir.caltech.edu}
\affiliation{Theoretical Astrophysics, California Institute of Technology, Pasadena, CA 91125, USA}
\affiliation{Kellogg Radiation Laboratory, California Institute of Technology, Pasadena, CA 91125, USA}
\affiliation{Department of Physics, School of Science, The University of Tokyo, Tokyo 113-0033, Japan}


\date{\today}

\begin{abstract}

We investigate contributions to the extragalactic gamma-ray background (EGB) due to neutralino dark matter (DM) 
pair-annihilation into photons, from DM density enhancements (minispikes) surrounding intermediate-mass black holes 
(IMBHs). We focus on two IMBH formation scenarios; our conservative scenario where IMBHs are remnants of Population-III 
stars, and our optimistic scenario here IMBHs are formed in protogalactic disks. In both scenarios, their formation in 
pregalactic halos at high redshift lead to the formation of minispikes that are bright sources of gamma-ray photons. 
Taking into account minispike depletion processes, we only sum contributions from a cosmological distribution 
of IMBHs with maintained minispikes. Our conservative scenario (BH mass $10^2 M_\odot$ with a $r^{-3/2}$ minispike) 
predicts gamma-ray fluxes that are an order larger than the equivalent flux, using the same DM parameters (mass 
$100\,\mathrm{GeV}$ and annihilation cross-section $3\times10^{-26}\,\mathrm{cm^3\,s^{-1}}$), from the host halo without 
IMBH minispikes. Our optimistic scenario (BH mass $10^5 M_\odot$ with a $r^{-7/3}$ minispike) predicts fluxes that are 
three orders larger, that can reach current EGB observations taken by EGRET (DM parameters as above). This fact may serve 
interesting consequences for constraining DM parameters and elucidating the true nature of IMBHs. Additionally, we determine 
the spectra of DM annihilation into monochromatic gamma-rays, and show that its flux can be within observational range of 
GLAST, providing a potential `smoking-gun' signature of DM. 
\end{abstract}


\maketitle


\section{\label{sec:intro}Introduction}

Despite compelling indirect evidence, from galactic to cosmological scales, the fundamental nature of the dominant 
non-baryonic component in the matter density of the universe (dark matter, hereafter DM) remains unknown. Intriguingly, 
extended models of particle physics independently provide us with a host of particle candidates for this as yet unknown 
matter, of which the most popular is the supersymmetric neutralino (see reviews \cite{JungmanKamionGriest,Bergstrom,
BertoneHooperSilk} for details). Upgrades of underground direct detectors looking for scattering of DM particles from 
nuclei, together with future neutrino, antimatter, and gamma-ray detectors looking for products of DM annihilation, will 
dramatically enhance our chances of understanding the true nature of DM. In particular, the forthcoming launch of the 
Gamma Ray Large Area Space Telescope (GLAST) \cite{GLAST} and numerous ground based Atmospheric Cerenkov Telescopes 
make indirect gamma-ray search especially promising.

Since the DM annihilation rate scales as the DM density squared, there is great advantage in observing areas where the 
DM density is believed to be high. The galactic centre (GC) is the immediate choice, and indeed strong gamma-ray emission 
has been observed and its nature and origin have been investigated by many researchers \cite{Bengtsson,Berezinsky,
BergstromUllio,BergstromUllioBuckley,BergstromEdsjoGunnarsson,Cesarini_etal,Hooper_etal,Fornengo_etal,Horns}. However, the 
DM density in the GC is highly uncertain, making accurate predictions difficult. For example, DM enhancements called `spikes' 
can form during the formation of a central supermassive-BH (SMBH) \cite{GondoloSilk}, but it can also be depleted by various 
processes by varying degrees \cite{UllioZhaoKamion,Merritt,Merritt_etal,BertoneMerritt}. In addition, nearby 
astrophysical (non-DM) gamma-ray sources make a potential DM detection impossible for all but a narrow range of DM parameters 
\cite{ZaharijasHooper}. 
Intermediate-mass black holes (IMBHs, see e.g.~\cite{MillerColbert}) provide an alternative source that may work positively
for DM detection. Bertone et al.~\cite{BertoneZentnerSilk} recently investigated the possibility of detecting a `smoking gun' 
gamma-ray signature of DM using IMBHs in the Milky-Way as point sources. They showed that IMBH formation increases the 
DM density in its vicinity to produce a `minispike', and also that DM enhancement depletion processes are generally less 
significant for IMBHs due to their roughly spherical distribution about the GC. They conclude that under optimistic 
circumstances, the Energetic Gamma Ray Experimental Telescope (EGRET) may have already seen a few of the IMBH minispikes 
as unidentified sources. 

Another avenue of indirect DM search is via the extragalactic gamma-ray background (hereafter EGB) measured over a wide 
energy range \cite{Sreekumar_etal,StrongMoskalenkoReimer}. The origin of this background is currently unknown, and it has 
been speculated that DM annihilation gamma-rays from cosmological distributions of DM contribute to some degree \cite{
BergstromEdsjoUllio,Ullio,TaylorSilk,ElsasserMannheim,ElsasserMannheim2,Ando,OdaTotaniNagashima,AndoKomatsu}. Since the 
DM annihilation cross-section is so small, a consideration of DM enhancements is crucial for meaningful gamma-ray flux 
predictions. A popular DM enhancement is those at the centres of galactic DM halos. However, Ando \cite{Ando} has recently 
shown that they are strongly constrained by observations of our galaxy. The author assumes universality of galactic DM halo 
profiles, and shows that DM annihilation cannot significantly contribute to the EGB without exceeding gamma-ray observations 
from our GC \cite{Ando}. The author also points out that this constraint could be loosened when one takes DM substructures
into account. 

In this paper we argue IMBHs minispikes as a \emph{substructure in the DM halo, that can lead to enhancements that do not 
conflict with current observations of our galaxy}. We determine contributions to the EGB from 
IMBH minispikes by summing gamma-ray fluxes from all redshifts. IMBH minispikes are not expected to greatly suffer 
from depletion processes, but we do take into account BH-BH mergers, which are known to strongly deplete minispikes and 
do occur in IMBHs. We also consider a conservative case ($10^2 M_\odot$ BHs of Population-III origin \cite{scenarioA} with
a $r^{-3/2}$ minispike) and an optimistic case ($\sim 10^5 M_\odot$ BHs formed in the centres of protogalactic disks \cite{
scenarioB} with a $r^{-7/3}$ minispike). Our result is that contributions to the EGB are increased by $1-3$ orders in magnitude. 
In particular, our optimistic case predicts fluxes that can reach current EGB observations. As this has interesting implications 
for constraining DM parameters and IMBH scenarios, we critically assess uncertainties in our calculation. We then determine 
the flux of DM annihilation into line gamma-rays, and show that under optimistic conditions, it is observable by GLAST. This 
provides a potential `smoking-gun' signature of DM.

This paper is structured as follows. In Sec.~\ref{sec:imbh} we introduce IMBHs, starting with their existence, followed by 
their formation scenarios, and finishing off with a summary of recent numerical studies. Then in Sec.~\ref{sec:formulations} 
we develop our calculation frameworks, first for the EGB, followed by DM annihilation, then minispike formation, moving 
finally on to our IMBH number density fitting. Calculations and results are in Sec.~\ref{sec:calc}, and discussions 
and conclusions in Sec.~\ref{sec:conc}. In all our calculations we adopt the standard flat cosmological constant plus cold DM 
($\Lambda$CDM) cosmology, with $\Omega_M = 0.3$, $\Omega_\Lambda = 0.7$, $h = 0.7$, and $\sigma_8=0.9$.

\section{\label{sec:imbh}IMBH: Evidence and Properties}

\subsection{\label{sec:evidence}Evidence for IMBH}

Clues for the existence of a class of BHs with masses heavier than stellar BHs but lighter than supermassive black holes 
(SMBH) have accumulated in recent years. We call these BHs intermediate-mass BHs (IMBHs) and loosely define their mass 
range as $20 \lesssim M_{bh} /M_\odot \lesssim 10^6$. 
We briefly discuss the observational evidences and theoretical motivations for their existence.

Observationally, studies of objects known as ultra-luminous x-ray sources (ULXs, \cite{Makishima_etal,Fabbiano}) reveal 
that they may harbour IMBHs. Although most x-ray sources can be understood as accretion by compact objects such as BHs, 
there is an upper limit on the luminosity for a given BH mass, known as the Eddington limit. For $20 M_\odot$, commonly
accepted as the upper limit of stellar BHs, this limit is $\sim 2.8\times 10^{39} \, \mathrm{ergs \, s^{-1}}$. ULXs are 
observed to exceed this limit. This phenomena can be explained by several mechanisms, including a short-term 
super-Eddington phase \cite{Begelman}, beaming \cite{King_etal}, or normal accretion by an IMBH. Although the debate has 
not been settled, evidence favouring the IMBH mechanism over the other two have accumulated in recent years; these come 
in various forms, including spectral analysis \cite{Mushotzky,MillerColbert}, evidence for a low temperature 
($\sim0.1\,\mathrm{keV}$) black-body component \cite{MillerFabianMiller}, analysis of break frequencies of the power 
density spectrum \cite{Cropper_etal}, and observation of broad Fe lines and quasi-periodic oscillations (QPO) \cite{
StrohmayerMushotzky}. It seems that at least a fraction of the ULXs, in particular the most luminous ones, are IMBHs. 
The BH mass inferred from ULX observations is of the order $\sim 10^3 M_\odot$ (assuming no beaming and accretion 
efficiency $\sim 10 \%$). Also, from their positions in the host galaxies we can deduce an upper limit of 
$\leq 10^6 M_\odot$, in order not to sink to the galactic centre by dynamical friction within a Hubble time \cite{
MillerColbert}. 

Theoretically, the existence of a population of IMBHs helps explain the origin of SMBHs. There is as yet no definitive 
SMBH formation scenario, but the discovery of quasars at redshifts $z<\sim6$ in the Sloan Digital Sky Survey \cite{
Fan_etal,Fan_etal2,Barth_etal,WillottMcLureJarvis,HaimanLoeb} suggests that they were already formed at high redshifts. 
Such an early formation lends itself to scenarios with massive seed-BHs, frequent mergers, and rapid accretion. A 
hierarchical formation scenario starting from massive seed-BHs also helps explain the tight correlations observed 
between the SMBH mass and properties of the host galaxy and halo \cite{scenarioB}. The natural outcome of this 
hierarchical scenario is the existence of `wandering' BHs in galactic halos, resulting from seed-BHs that did not 
successfully merge into a SMBH \cite{IslamTaylorSilk,KoushiappasZentner,VolonteriHaardtMadau}, which we call IMBHs. 
However, despite the theoretical interests, it is difficult to obtain conclusive evidence for their existence. One 
viable strategy is to search for gravitational-waves produced in seed-BH mergers, which may become possible with the 
launch of the Laser Interferometer Space Antenna (LISA) \cite{LISAobs,KoushiappasZentner}.

\subsection{\label{sec:formation}IMBH Formation Scenarios}

In this study, we follow a previous study by Bertone et al.~\cite{BertoneZentnerSilk} and focus on two seed-BH formation 
scenarios covering the wide range of possible IMBH masses. In the first scenario, which we refer to as scenario A, the 
seed-BHs are remnants of the collapse of Population-III (or first generation) stars \cite{scenarioA}. These stars are 
generally massive due to suppression of mass-loss processes, a result of their very low-metal composition. As such, they 
are also called zero-metalicity or very massive stars (VMS). The fate of VMS have been studied by e.g.~\cite{BondArnettCarr,
FryerWoosleyHeger}: stars with $100 \lesssim M/M_\odot \lesssim 250$ encounter the electron-positron pair instability and 
explode in a giant nuclear-powered explosion, leaving no compact remnant, while heavier stars collapse completely to a BH 
containing at least half of the stellar mass \cite{BondArnettCarr}.

What we need to know is the mass-function of these BHs. Unfortunately, the mass-function of Population-III stars is not 
well known, although recent studies indicate a double peaked function that extends up to a few $10^3 M_\odot$ \cite{
NakamuraUemura}. We base our scenario on the one proposed in \cite{scenarioA} and further studied by \cite{ZhaoSilk,
BertoneZentnerSilk,IslamTaylorSilk,VolonteriHaardtMadau}. Interestingly, if BHs with masses $\gtrsim 10^2 M_\odot$ form 
at high redshifts in minihalos representing $\sim 3 \sigma$ peaks of the smoothed density field, the resulting baryon 
mass fraction is found to be comparable to those observed for SMBHs in nearby galaxies \cite{scenarioA}. As in \cite{
BertoneZentnerSilk}, we conservatively consider BHs of mass $10^2 M_\odot$ forming in minihalos of masses larger than 
$M_{v,crit} \sim 10^7 M_\odot$ at formation redshift $z_f \sim 18$. 

In the second scenario, which we refer to as scenario B, BHs form from low angular momentum gas in protogalactic disks at 
high redshifts, producing a population of seed-BHs with masses $\sim 10^5 M_\odot$. We use the scenario proposed in \cite{
scenarioB}, which we briefly summarise. During the collapse of the first halos, gas cools and a pressure supported disk 
forms if the halo is massive enough to contain a relatively large fraction of molecular hydrogen (molecular hydrogen is the 
main coolant, see \cite{hydrogencooling}). Local gravitational instabilities in the disk manifest themselves as an effective 
viscosity that transfers angular momentum outwards and cause an inflow of gas, in particular the low angular momentum tail. 
In halos that are both massive enough to contain enough hydrogen for cooling \emph{and} do not experience mergers with other 
halos, the protogalactic disk can evolve uninterrupted until ultimately being terminated by the heating and disruption caused 
by supernovae of population-III stars. During this time, mass transfer of order $\sim 10^5 M_\odot$ occurs. The central mass 
may be briefly pressure supported but will ultimately collapse to a BH due to post-Newtonian instabilities. Since population-III 
stars have typical lifetimes of $1 - 10 \,\mathrm{Myrs}$, the halo must be avoid of mergers over many dynamical times. This sets 
a stringent lower limit on the required halo mass. At $z_f=12$, this limit is $M_{v,crit}=10^8 M_\odot$ (see \cite{scenarioB} 
for an expression). The masses of the BHs that form have a near log-normal distribution with peak mass 
$M_{bh,0}=2.3 \times 10^5 M_\odot$ independent of $z_f$, with spread $\sigma_{bh}=0.9$ \cite{scenarioB}. 


Scenario B naturally contains many parameters other than $z_f$ (such as fraction of gas cooled, lifetime of population-III 
stars, etc). However, uncertainties in these parameters largely affect $M_{bh,0}$, and as such are ultimately masked by 
uncertainties in $z_f$, which affects the \emph{total number} of seed-BHs formed. We therefore treat scenario B formation 
to be described by one parameter, $z_f$. Now, the epoch of cosmological reionisation places a lower limit on $z_f$. This is 
because the heating of the intergalactic medium and the subsequent ionisation of molecular hydrogen (i.e.~reionisation) 
terminates further baryon cooling. Without molecular hydrogen, scenario B formation cannot proceed, even in the heaviest 
halos which satisfy $M_{v}>M_{v,crit}$. Therefore, for scenario B, $z_f=z_{re}$, where $z_{re}$ is the redshift of reionisation 
(we assume for simplicity that seed-BH formation stops abruptly at reionisation).

\subsection{\label{sec:simulation}IMBH Number Density: Results from Numerical Studies}

Now that we have discussed IMBH formation, we will summarise how their number density evolves with time. Since in this work 
we are only interested in IMBHs with minispikes, from here on we will use `IMBH' to imply `an IMBH with a maintained minispike'. 
First, although seed-BH formation can lead to minispike formation, various processes destroy it by varying degrees (see 
Sec.~\ref{sec:minispike}), and the strength and number of minispike decreases with time. BH mergers, i.e.~mergering with 
another IMBH or a SMBH, are the most destructive and we must therefore consider its effects. Before doing so however (in 
Sec.~\ref{sec:evolution}) we summarise basic procedures and results of previous numerical studies.

The basic approach focuses on constructing a statistical sample of halo formation histories (each called a `realisation') 
in the context of the hierarchical CDM model for structure formation, followed by computing the dynamical evolution of halos 
and BHs within halos (see \cite{BertoneZentnerSilk,KoushiappasZentner,Zentner_etal,IslamTaylorSilk} for further details). The 
first step is to consider a virialised halo of mass $M_{v,0}$ at $z=0$ and construct a merger history, i.e.~a list of all the 
smaller halos that merged together to form the final halo, as well as the redshifts at which the mergers occurred. The next 
step is to plant BHs in halos that satisfy seed-BH formation conditions. That is, if a halo satisfies $M_v>M_{v,crit}$ and 
$z>z_f$, a seed-BH is planted at the earliest time $M_v>M_{v,crit}$ still holds. This point is labelled $z_{bh}$. The last 
step involves evolving the halos and BHs forwards to $z=0$, as described in \cite{KoushiappasZentner,Zentner_etal}. During 
the last step, a BH that comes within a distance $\mathrm{min}(0.01r_{vir},1\mathrm{kpc})$, where $r_{vir}$ is the virial 
radius of the halo, of another BH is considered a BH-BH `potential pair', and is decoupled from the simulation. Previous 
work focused on the gravitational-waves produced by the potential pairs' subsequent mergers \cite{KoushiappasZentner}, but 
for our purposes we exclude all `potential-pairs' from our EGB calculation.

What necessary results can we obtain from these studies? First, the distribution of BH formation redshifts $z_{bh}$ 
is found to be exponential, i.e.~using the Milky-way galaxy with $M_{v,0} = 10^{12.1} h^{-1} M_\odot$ and 200 realisations, 
\cite{KoushiappasZentner} finds that the number of BHs formed peaks at $z_f$ and decreases exponentially for higher $z_{bh}$ 
(see their Fig.~2). 
Second, by following the dynamical evolution of BHs, \cite{BertoneZentnerSilk} finds that $N_{bh,A}=1027 \pm 84$ 
scenario A BHs and $N_{bh,B}=101 \pm 22$ scenario B BHs remain unmerged in a Milk-way sized halo. Errors denote the $1\sigma$ 
halo-to-halo scatter (200 realisations were performed). Finally, the number of `potential pairs' formed is highest at $z_f$ 
when the BH number density is largest, and decreases as a power-law of $(1+z)$ with decreasing z (see their Fig.~3).

\section{\label{sec:formulations}Formulations}

\subsection{\label{sec:egb}Extragalactic Gamma-Ray Background}

To calculate contributions to the EGB flux from unresolved cosmological DM sources, we adopt the methodology in Ullio et 
al.~\cite{Ullio}, but extend it for our purposes of including IMBH minispike enhancements.

\newcommand{\ud}{\mathrm{d}}

Let $\ud \mathcal{N}_\gamma/ \ud E(E,M_{bh},z)$ be the differential energy spectrum for the number of gamma-ray 
photons emitted per unit time from a single IMBH of mass $M_{bh}$ at redshift $z$, and $\ud n/ \ud M_{bh}(M_{bh},z)$ 
the comoving number density of IMBHs of mass $M_{bh}$ at redshift $z$. From these we can determine the number of photons 
emitted in a proper volume $\ud V$ say, at redshift $z$ in time interval $\ud t$ and energy range $(E,E + \ud E)$. Then, 
assuming isotropic emission, the corresponding number of photons $\ud N_\gamma$ collected by a detector on Earth with an 
effective area $\ud A$ in the redshifted energy range $(E_0,E_0 + \ud E_0)$ over a time $\ud t_0$, is
\begin{eqnarray} \label{gamma_earth} 
\ud N_\gamma &=& \int \ud M_{bh} (1+z)^3 \frac{\ud n}{\ud M_{bh}}(M_{bh},z) 
\frac{\ud \mathcal{N}_\gamma}{\ud E}(E,M_{bh},z) \nonumber \\
& & \times \; e^{-\tau(E_0,z)}
\frac{\ud V \ud A}{4 \pi (R_0 r)^2} \ud E_0 \, \ud t_0, 
\end{eqnarray}
where we have used the fact that $\ud E \ud t=(1+z)\ud E_0(1+z)^{-1}\ud t_0=\ud E_0 \ud t_0$. The exponential term is an 
attenuation factor which accounts for the absorption of gamma-rays during propagation to Earth, the factor $(1+z)^3$ 
converts from comoving to proper IMBH number density, and $R_0$ comes from the metric of our cosmology as defined in 
section \ref{sec:intro}. We define $\ud V$ by radial increment $\ud r$ and angular increment $\ud \Omega$ as 
\[
\ud V = \frac{(R_0 r)^2 R_0}{(1+z)^3} \ud r \ud \Omega.
\]

The main absorption of gamma-rays of a few GeV is via pair production on the extragalactic background light emitted by 
galaxies in the optical and infrared bands. Although at these gamma-ray energies attenuation is almost negligible, we 
use the form $e^{-z/z_{max}}$, where we approximate $z_{max} \simeq 3.3 (E_0/10 \,\mathrm{GeV})^{-0.8}$, which is a 
parametrisation that reproduces results of \cite{SalamonStecker} with enough accuracy. Substituting $\ud V$ into 
Eq.~(\ref{gamma_earth}) and changing the integration along the line of sight $\ud r$ to along $\ud z$, we get for the flux 
\begin{eqnarray} \label{flux} 
\frac{\ud \Phi_\gamma}{\ud E_0} &\equiv&
\frac{\ud N_\gamma}{\ud A \ud \Omega \ud t_0 \ud E_0} \nonumber \\
&=& \frac{c}{4\pi} \int \ud z 
\frac{e^{-z/z_{max}}}{H_0 h(z)} 
\int \ud M_{bh} \frac{\ud n}{\ud M_{bh}}(M_{bh},z)  \nonumber \\
& & \times \frac{\ud \mathcal{N}_\gamma}{\ud E}(E_0(1+z),M_{bh},z) ,
\end{eqnarray}
where $c$ is the speed of light, $H_0$ is the Hubble parameter now, and $h(z)$ is the function 
$h(z)=\surd[\Omega_M(1+z)^3+\Omega_\Lambda]$. 

We use Eq.~(\ref{flux}) to compute the final EGB flux. In the following subsections, we will deal with the physical 
quantities in the expression in detail. 

\subsection{\label{sec:annihilation}Dark Matter Annihilation Gamma-rays}

In order for DM particles to satisfy cosmological constraints, they are expected to have a small but non-zero annihilation 
cross-section into Standard-Model particles. This ensures they are in chemical equilibrium in the early universe. By 
constraining their relic density by cosmological observations, one can obtain limits on their annihilation cross-section. 
For DM that is a thermal relic, the required cross-section is $3 \times 10^{-26} \,\mathrm{cm^3\,s^{-1}}$. However this 
should be taken as an upper limit, because processes such as coannihilation can allow smaller cross-sections (see \cite{
JungmanKamionGriest,BertoneHooperSilk} for details). The DM particle mass is constrained from below by collider experiments 
and above by theory, giving a commonly accepted range of $m_\chi \sim 50\,\mathrm{GeV}-10\,\mathrm{TeV}$.

Now, the flux of gamma-rays from a single IMBH minispike is quantified by the term $\ud \mathcal{N}_\gamma/ \ud E(E,M_{bh},z)$ 
in Eq.~(\ref{flux}). For gamma-rays of DM origin, we can rewrite
\begin{eqnarray} \label{gamma_source}
\frac{\ud \mathcal{N}_\gamma}{\ud E} &=& \frac{\sigma v}{2} 
\frac{\ud N_\gamma(E)}{\ud E} \int^{r_{sp}}_{r_{lim}} n_\chi^2 \ud^3r \nonumber \\
&=& \frac{\sigma v}{2} \frac{\ud N_\gamma(E)}{\ud E} 
\frac{1}{m_\chi^2} \int^{r_{sp}}_{r_{lim}} \rho^2(r) \ud^3r,
\end{eqnarray}
where the factor $1/2$ appears due to the fact this is an annihilation of identical particles \cite{Ullio}, $\rho(r)$ is the 
DM density profile around an IMBH ranging from radii $r_{lim}$ to $r_{sp}$ (we discuss these in Sec.~\ref{sec:minispike}), 
$\sigma v$ is the annihilation cross-section times relative velocity, and $\ud N_\gamma(E) / \ud E$ is the differential 
gamma-ray yield per annihilation. The latter can be divided into continuum and monochromatic (line) emissions, and can be 
written as \cite{Ullio}
\begin{eqnarray} \label{annihilationspectrum}
\frac{\ud N_\gamma(E)}{\ud E} &=&  \sum_Y b_{\gamma Y}n_{\gamma Y}\delta(E-m_\chi(1-M_Y^2/4m_\chi^2)) \nonumber \\
&& + \sum_F b_F \frac{\ud N_{cont}^F}{\ud E}(E),
\end{eqnarray}
where the second term refers to the continuum. The continuum photons are produced by annihilation 
into the full set of tree-level final states F including fermions and gauge or Higgs bosons which generate photons on decay. 
The bulk of the photons however are produced in the hadronisation and decay of neutral pions (decay mode $\pi^0 \to 2 \gamma$), 
with a branching ratio of $b_F=98.8\%$. The monochromatic emission on the other hand is the result of prompt annihilation into 
two-body states including a photon, a process that is forbidden at tree-level and only allowed in higher order perturbation 
theory. Although subdominant, these gamma-rays have the advantage of producing a `smoking gun' signature of DM annihilation,
i.e.~photons of energy $E=m_\chi$ for the $2\gamma$ final state and $E=m_\chi(1-M_Y^2/4m_\chi^2)$ for the photon and particle 
Y final state. The parameter $b_{\gamma Y}$ is the branching ratio into the respective channels, and $n_{\gamma Y}$ is the 
number of photons emitted per annihilation. 

In the current work we consider both the continuum and monochromatic gamma-ray emissions. For the continuum we consider the 
$\pi^0$ branch, and use a conveniently parametrised form of the rest frame energy distribution per annihilation, 
$\ud N_{cont}(E)/ \ud E = (0.42/m_\chi)e^{-8x}/(x^{1.5}+0.00014)$ where $x \equiv E/m_\chi$. For the monochromatic emission 
we consider the $\chi\overline{\chi} \to 2 \gamma$ process, which exists in many supersymmetric models. We use for the 
branching ratio $b_{2\gamma}=10^{-3}$. 

\subsection{\label{sec:minispike}Dark Matter Enhancement around IMBH}

The emergence of a deeper gravitational potential due to the formation of a BH inevitably alters the DM halo in which it is 
formed. It has been shown that the adiabatic formation of a SMBH results in an enhancement of the nearby DM density, called a 
`spike' \cite{GondoloSilk}. However, it has also been shown that spike formation depends on initial conditions \cite{
UllioZhaoKamion}, and even formed, it is depleted by processes such as BH-BH mergers \cite{Merritt_etal}, dynamical processes 
such as gravitational scattering off stars \cite{Merritt}, and DM annihilation itself \cite{BertoneMerritt}. Fortunately, 
minispikes around IMBHs may not be greatly affected by these problems. Firstly, we have selected IMBHs that survive without 
experiencing any major mergers. Secondly, these unmerged IMBHs are not necessarily localised in the galactic centre, and 
null observations of ULXs in our galaxy suggest they reside in satellite halos with no significant stellar component. These 
imply that dynamical processes are not very significant. Thirdly, a central BH formation is a built-in property of scenario B, 
predicting a strong minispike. However, we must make clear that the precise likelihood of minispike formation and survival 
are still uncertain. With these in mind, we treat scenario B as our `optimistic' case and scenario A our `conservative'. 

The BH in scenario B forms in the centre of its host halo \cite{scenarioB}, and therefore predicts a strong minispike. To 
compute its shape we need to specify the initial DM profile. If we write the initial inner DM profile as 
$\rho(r) \propto r^{-\gamma}$, the minispike has the form \cite{BertoneZentnerSilk}
\begin{equation} \label{spikeequation}
\rho_{sp}(r) = \rho(r_{sp}) \left( \frac{r}{r_{sp}} \right)^{-\gamma_{sp}},
\end{equation}
where $r_{sp}$ is the radius of the minispike and $\gamma_{sp}$ is the gradient, given as
\begin{equation} \label{spikegradient}
\gamma_{sp} = \frac{9-2\gamma}{4-\gamma}.
\end{equation}
Values for $\gamma$ have been proposed by analytic fits of \emph{N}-body simulations, e.g.~$\gamma=1$ by Navarro, Frenk 
and White (NFW) \cite{NFW}, and $\gamma=1.5$ by Moore et al.~\cite{MooreProfile}. Recent simulations have produced 
shallower profiles (see \cite{MerrittGrahamMoore} and references therein) and as well as somewhere between 1--1.5 \cite{
Navarro_etal,DiemandMooreStadel,FukushigeKawaiMakino}. In this work we assume a middle value of $\gamma=1$, but keep in 
mind that a smaller value will lead to less gamma-rays. For reference, our values for halo radii are summarised in Table 
\ref{table:radii}.

For scenario A, while some works show that they are formed in the centres of their host halos \cite{AbelBryanNorman}, others 
show fragmentation may lead to off-centre BHs \cite{BrommCoppiLarson}. Although motivations for a central 
BH are strong, to remain conservative we consider an off-centre BH formation. It has been shown that BH formation in an 
uniform DM background will form a mild $r^{-3/2}$ minispike \cite{ZhaoSilk,BertoneZentnerSilk,QuinlanHernquistSigurdsson,
Peebles}. Specifically, we consider the minispike studied in \cite{BertoneZentnerSilk,ZhaoSilk}, with $\rho(r)=\rho_h(r/r_h)^{-3/2}$ 
where $r_h=0.045\,\mathrm{pc}$.

In both scenario A and B minispikes, the very DM annihilation we are studying sets an lower limit on the minispike radius. 
Assuming DM annihilation is the main process by which the inner density decreases, the DM number density $n_\chi$ obeys the 
evolution equation $\dot{n}_\chi(r,t) = - \sigma v n_\chi^2(r,t)$. Solving this yields the solution
\begin{equation}
n_\chi(r,t) = \frac{n_\chi(r,t_f)}{1+n_\chi(r,t_f) \sigma v (t-t_f)},
\end{equation}
where $t-t_f$ is the time elapsed since BH formation. The upper limit is of order $m_{\chi}/\sigma v(t-t_f)$, and we define 
$r_{lim}$ as the radius at which the following holds
\begin{equation} \label{rlim}
\rho_{sp}(r_{lim}) = \frac{m_\chi }{\sigma v (t-t_f)} \equiv \rho_{lim},
\end{equation}
where $m_\chi$ is the DM mass. For common DM parameters, $r_{lim}$ has grown to $\simeq 10^{-3} \,\mathrm{pc}$ by $z=0$. 


\begin{table}
\caption{\label{table:radii}Representative values of halo radii for scenario B, for two formation redshifts. Shown are the halo 
virial radius $r_{vir}$, the scale radius $r_s$ which defines the shape of the NFW profile (see \cite{NFW}), the `influence' 
radius $r_h$ of the BH as defined in \cite{Merritt_book}, and the inner radius of the minispike $r_{lim}$. DM parameters are
$m_\chi=100\,\mathrm{GeV}$ and $\sigma\mathit{v}=3 \times 10^{-26} \,\mathrm{cm^3\,s^{-1}}$.}
\begin{ruledtabular}
\begin{tabular}{ccc}
 & Radii, $z_{f}=10.9$ [pc] & Radii, $z_{f}=12$ [pc] \\
\hline
$r_{vir}$ & 1300 & 1100 \\
$r_s$ & 510 & 480 \\
$r_h$ & 34 & 33 \\
$r_{sp}$ & 6.8 & 6.7 \\
$r_{lim}(z=0)$ & $5 \times 10^{-3}$ & $4 \times 10^{-3}$ \\
\end{tabular}
\end{ruledtabular}
\end{table}

\subsection{\label{sec:evolution}Modelling the IMBH Number Density}

Now we parametrise the decreasing IMBH number density, using results of previous numerical studies. To determine the initial 
(i.e.~at $z_f$) IMBH number density, we consider a delta-function seed-BH formation occurring at $z_f$, and plant BHs in 
halos that satisfy $M_v \geq M_{v,crit}$. This simplified picture neglects BHs that would have formed earlier. That is, for 
some halos we plant one BH where in fact there would be two (or more). However, as \cite{KoushiappasZentner} shows, the 
number of BHs formed in $z>z_f$ decreases exponentially. This, coupled to the fact that massive halos become increasingly 
unlikely for higher redshifts, the effect of ignoring them is minimal. We also stress that this picture in no way overestimates 
EGB contributions, because (i) we have underestimated the initial BH density, and (ii) we have neglected all DM annihilations 
before $z_f$. Moreover, any underestimation is masked by uncertainties caused by $z_f$. 

The required calculation is
\begin{equation} \label{nbh_zf}
n_{bh}(z_f)=\int_{M_{v,crit}}^{\infty} \ud M_v \frac{\ud n}{\ud M_v}(z=z_f),
\end{equation}
where $n_{bh}$ is the comoving number density of IMBHs that have not experienced any mergers, and $\ud n / \ud M_v$ is the 
halo mass-function. For the mass-function we use one postulated by Press-Schechter theory \cite{PressSchechter}, 
\begin{equation} \label{massfunction}
\frac{\ud n}{\ud M}=\frac{\overline{\rho}_0}{M^2}
\nu f(\nu)
\frac{\ud \log \nu}{\ud \log M},
\end{equation}
where $\overline{\rho}_0$ is the comoving matter background density, $\overline{\rho}_0 = \rho_c \Omega_M$, with $\rho_c$ 
the critical density. The parameter $\nu \equiv \delta_{sc}(z)/\sigma(M)$ is defined as the ratio between the critical 
linear fractional overdensity required for vitalisation over $\sigma(M)$ the present rms linear density fluctuation in spheres 
containing a mean mass $M$. For the multiplicity function $f(\nu)$ we use the ellipsoidal collapse model \cite{ShethTormen}, 
normalised to results of \emph{N}-body simulations of the Virgo Consortium \cite{Jenkins_etal}, as was done in \cite{Ullio}. 

Now, we assume that the number of IMBHs present and the number of `potential pairs' formed are proportional, and fit a 
power-law redshift dependency to $n_{bh}$,
\begin{equation} \label{IMBHdensity}
n_{bh}(z)=n_{bh}(z_f) 
\left( \frac{1+z}{1+z_f} \right)^\beta,
\end{equation}
where $\beta$ is a free parameter. We can derive representative values for $\beta$ by fitting $n_{bh}(0)$ to numerical 
results by \cite{BertoneZentnerSilk}. From their value of $N_{bh,A}$ ($N_{bh,B}$), we interpolate the number of unmerged
IMBHs in other galaxies by their halo masses, and determine $n_{bh}(0)$ as
\begin{equation} \label{nbh_0}
n_{bh}(0) = N_{bh} \int_{M_{v,crit}}^{\infty}
\ud M_v 
\left( \frac{M_v}{10^{12.1} h^{-1} M_\odot} \right)^\alpha
\frac{\ud n}{\ud M_v}(0),
\end{equation}
where $\alpha$ is some constant which we assume to be 1. This gives $\beta \simeq 0.2$ ($0.3$) for scenarios A 
(scenario B). Although in reality we do not expect $\alpha$ to be exactly 1 (e.g.~we expect less unmerged IMBHs in older 
elliptical galaxies), we find that for reasonable deviations from $\alpha=1$ our estimates of $\beta$ do not appreciably 
change. Moreover, uncertainties caused by deviations in $\alpha$ are far smaller than those caused by $z_f$. 

Lastly, we conclude this section with the parameters of scenarios A and B in Table \ref{table:parameters}.


\begin{table}
\caption{\label{table:parameters}Scenario A and B parameters, including formation redshift $z_f$, the 
minimum halo mass required for formation to occur $M_{v,crit}$, the number density of IMBH $n_{bh}$ at $z=z_f$ determined from
Eq.~(\ref{nbh_zf}), the number of IMBHs residing in a Milky-Way size galactic halo $N_{bh}$ \cite{BertoneZentnerSilk}, $n_{bh}$ 
at $z=0$ derived from Eq.~(\ref{nbh_0}), and $\beta$ derived from our fitting.}
\begin{ruledtabular}
\begin{tabular}{ccc}
 & Scenario A & Scenario B \\
\hline
$z_f$ & 18 & $10.9^{+2.3}_{-2.7}$ \footnote{$z_{re}$ taken from WMAP 3rd year results \cite{Page_etal}} \\
$M_{bh}$ $[M_\odot]$ & $10^2$ & $10^5$ \\
$M_{v,crit}(z_f)$ $[M_\odot]$ & $4 \times 10^6$ & $10^8$ \\
$n_{bh}(z_f)$ $[\mathrm{Mpc^{-3}}]$ & 23 & 2.5 \\
$N_{bh}$ & $1027\pm84$ & $101\pm22$ \\
$n_{bh}(0)$ $[\mathrm{Mpc^{-3}}]$ & 12 & 1.1 \\
$\beta$ & 0.2 & 0.3 \\
\end{tabular}
\end{ruledtabular}
\end{table}

\section{\label{sec:calc}Calculation and Results} 

\subsection{Contribution to the EGB}

Substituting Eq.~(\ref{gamma_source}) into Eq.~(\ref{flux}) we find the gamma-ray flux observed at Earth as
\begin{equation} \label{finalflux}
\frac{\ud \Phi_\gamma}{\ud E_0} =
\frac{\sigma v}{8\pi} \frac{c}{H_0} 
\frac{\overline{\rho}_0^2}{m_\chi^2} \int^{z_f}_0 \ud z 
\frac{\Delta^2(z)}{h(z)}
\frac{\ud N_\gamma(E')}{dE}
e^{-z/z_{max}},
\end{equation}
where $E'=E_0(1+z)$ and $\Delta^2(z)$ is defined
\begin{equation} \label{delta}
\Delta^2(z) = \frac{1}{\overline{\rho}_0^2}
\int \ud M_{bh} \frac{\ud n}{\ud M_{bh}}(M_{bh},z)
\int^{r_{sp}}_{r_{lim}} \rho^2(r) \ud^3r.
\end{equation}
Note the disappearance of the $(1+z)^3$ term in Eq.~(\ref{finalflux}) compared to \cite{Ullio}. The reason for this is that 
while the DM density in the halo is a function of $z$, the density in the minispike around an IMBH is only a function of $z_f$; 
the $z$ dependence is taken into account instead by $r_{lim}$. 

For scenario B, since the BH mass distribution is near log-normal, and the enhancement due to the minispike scales linearly 
with the BH mass, we find that with very good accuracy we can approximate all the BH mass to have the peak value $M_{bh,0}$. 
Hence, we substitute $\ud n/\ud M_{bh}(M_{bh},z) = n_{bh}(z)\delta(M_{bh}-M_{bh,0})$, which gives the final form we will use,
\begin{equation} \label{delta2}
\Delta^2(z) = \frac{1}{\overline{\rho}_0^2}
n_{bh}(z) \int_{r_{lim}}^{r_{sp}} \rho(r)^2 \ud^3r.
\end{equation}

\begin{figure}
\includegraphics[scale=0.4]{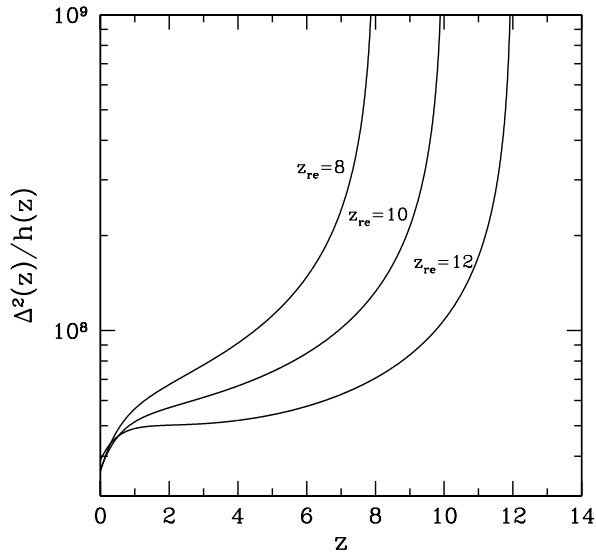}
\caption{\label{fig:enhancement}Enhancement factor $\Delta^2$ as a function of redshift, for scenario B IMBH minispikes. Three 
values of $z_f=z_{re}$ are plotted, 8, 10, and 12. The peak enhancement at $z_f$ is largely due to the fact that the minispike 
is sharpest just after it is formed. This demonstrates the effectiveness of the minispike in enhancing contributions to the 
EGB.}
\end{figure}

In Fig.~\ref{fig:enhancement} we show the redshift dependence of the enhancement function $\Delta(z)^2 / h(z)$ for scenario B, 
for three values of $z_f=z_{re}=$ 8, 10, and 12. The strongest enhancement occurs at $z_f$, partly because of the highest IMBH 
number density, but mainly because of the presence of the sharpest minispike. This displays the merits of considering the EGB, 
because high redshift enhancements are most easily observed as a diffuse background. On the other hand, we must therefore 
carefully consider the IMBH number density at $z_f$. However, a major uncertainty in the hierarchical formation scenario of SMBHs 
is that the `occupation number', that is, the fraction of galaxies containing a seed-BH at high redshift, is highly model dependent. 
In other words, contributions to the EGB flux can potentially constrain seed-BH models. To this effect, our calculation for the 
initial density is a conservative value (see Sec.~\ref{sec:evolution}).

\begin{figure}
\includegraphics[scale=0.4]{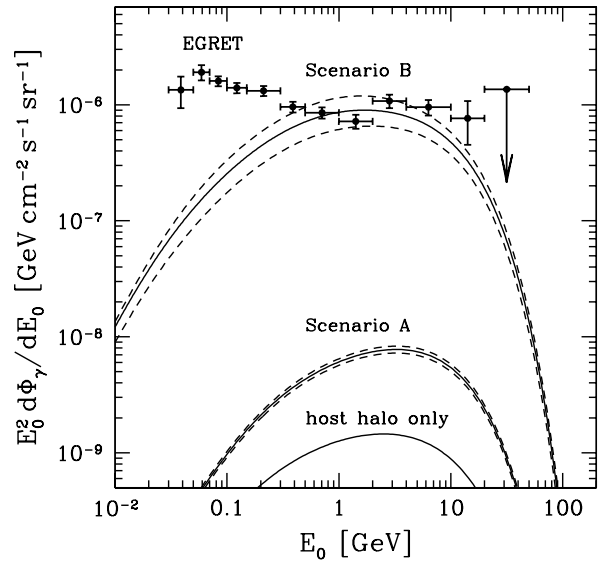}
\caption{\label{fig:flux}Contributions to the EGB flux, from scenario A and scenario B IMBH minispikes. Also shown are EGRET 
data and predictions of our minimal host halo only scenario (i.e.~no spikes and minispikes). We see that minispikes increase EGB 
contributions by 1 -- 3 orders in magnitude. The $1\sigma$ scatter in $N_{bh,A}$ \cite{BertoneZentnerSilk} is shown for scenario A. 
For scenario B, the $1\sigma$ scatter in $N_{bh,B}$ as well as $z_{re}$ are shown combined. In all calculations, $z_f$ and $\beta$ 
shown in Table \ref{table:parameters} are used, with DM parameters $m_\chi=100 \,\mathrm{GeV}$ and 
$\sigma\mathit{v}=3 \times 10^{-26} \,\mathrm{cm^3\,s^{-1}}$. Note that scenario A is our conservative case; a BH of mass 
$10^2M_\odot$ with a mild $r^{-3/2}$ minispike.}
\end{figure}

In Fig.~\ref{fig:flux} we show contributions to the EGB from DM annihilation into continuous gamma-rays (we discuss monochromatic 
emission later). We use DM parameters $m_\chi=100 \,\mathrm{GeV}$ and $\sigma v=3 \times 10^{-26} \,\mathrm{cm^3\,s^{-1}}$. The 
values of $z_f$ and $\beta$ in Table \ref{table:parameters} were used.
We do not expect low-redshift IMBHs to be resolved; it has been shown that GLAST will resolve minispikes in our galaxy and 
possibly Andromeda, but not further \cite{BertoneZentnerSilk}. Moreover, we find that sources within $z=0.01$ contribute less 
than a tenth of the total EGB contribution. We therefore show the flux, integrated safely from $z=0$ to $z_f$. We find that 
minispikes increase the gamma-ray flux from DM halos by $1-3$ orders. In particular, scenario B can give fluxes that are of 
the order of current observations. Scenario A fluxes are two orders smaller, but we stress that scenario A is a conservative 
case, using the smallest IMBH mass and a mild $r^{-3/2}$ minispike. The `host halo only' is a prediction without any spikes 
nor minispikes, and therefore acts as a minimal prediction. The dashed curves indicate uncertainties caused by $1\sigma$ scatter 
in: $N_{bh,A}$ for scenario A, and $N_{bh,B}$ and $z_{re}$ combined for scenario B.

\begin{figure}
\includegraphics[scale=0.4]{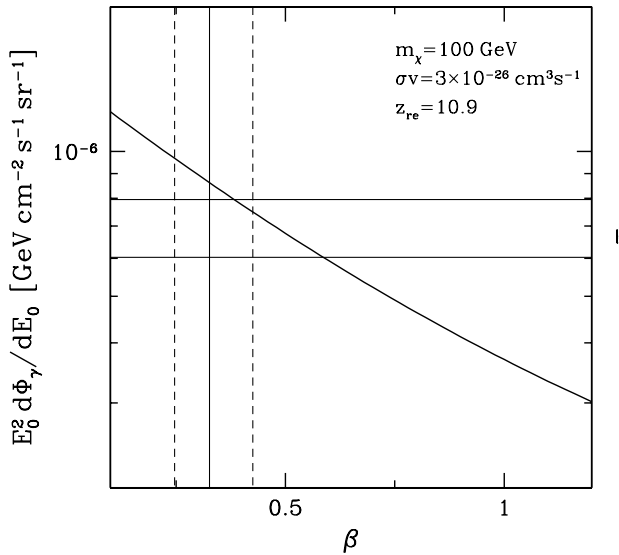}
\caption{\label{fig:beta}Plot showing how EGB contributions from scenario B IMBH minispikes varies due $\beta$, the parameter 
of our IMBH number density fitting. Plotted is the flux at an energy of 1 GeV, where the contribution is greatest. The range 
of $\beta$ determined from the $1\sigma$ scatter in $N_{bh,B}$ is shown by the vertical dashed lines. EGRET observations are 
shown by the horizontal rectangle.}
\end{figure}

In Fig.~\ref{fig:beta} we show how EGB contribution from scenario B depends on the free parameter $\beta$. As before, we use 
$z_{re}=10.9$, $m_\chi=100 \,\mathrm{GeV}$, and $\sigma v=3 \times 10^{-26} \,\mathrm{cm^3\,s^{-1}}$. We show the $1\sigma$ 
scatter in $N_{bh,B}$ using vertical dashed lines. We find that our fitting is within $1\sigma$ error of EGRET observations. 

\subsection{Constraining DM Parameters}

\begin{figure}
\includegraphics[scale=0.4]{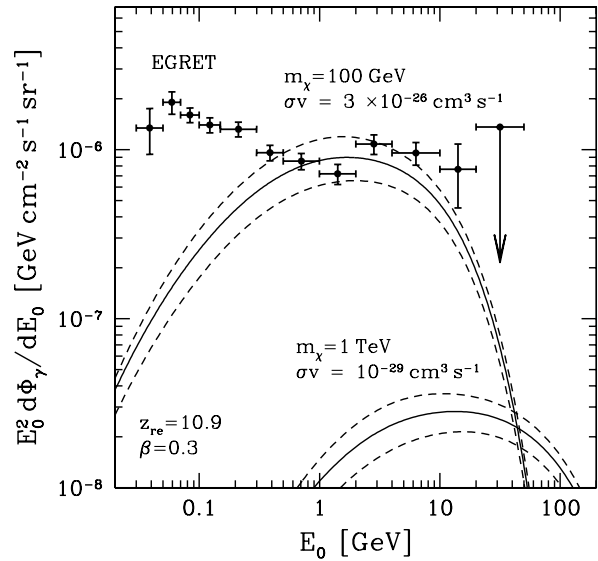}
\caption{\label{fig:cases}Contributions to the EGB from scenario B IMBH minispikes, for two sets of DM parameters, the first 
with $m_\chi=100 \,\mathrm{GeV}$ and $\sigma\mathit{v}=3 \times 10^{-26} \,\mathrm{cm^3\,s^{-1}}$ and the second with 
$m_\chi=1 \,\mathrm{TeV}$ and $\sigma\mathit{v}=10^{-29} \,\mathrm{cm^3\,s^{-1}}$. Errors include those from $N_{bh,B}$ and 
$z_{re}$.}
\end{figure}

Instead of performing a complete scan over DM parameters space, we follow Bertone et al.~\cite{BertoneZentnerSilk} and consider 
two discrete cases. In addition to our previously chosen set 
$m_\chi=100 \,\mathrm{GeV}$ and $\sigma\mathit{v}=3 \times 10^{-26} \,\mathrm{cm^3\,s^{-1}}$, we define a new set 
$m_\chi=1 \,\mathrm{TeV}$ and $\sigma\mathit{v}=10^{-29} \,\mathrm{cm^3\,s^{-1}}$. 
Using scenario B with $z_{re}=10.9$ and $\beta=0.3$, we plot in Fig.~\ref{fig:cases} the predicted EGB flux for both sets.

\begin{figure}
\includegraphics[scale=0.4]{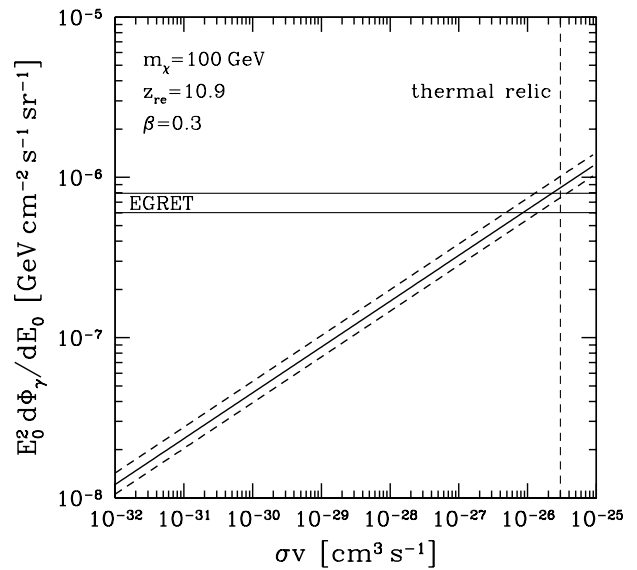}
\caption{\label{fig:crosssec}Gamma-ray flux at $E=1\,\mathrm{GeV}$ from scenario B IMBHs, plotted against the annihilation 
cross-section $\sigma\mathit{v}$. The dashed lines indicate combined error in $N_{bh,B}$ and $z_{re}$. EGB observations by 
EGRET and $\sigma v$ of `natural' thermal neutralinos are shown. The weak dependency on $\sigma v$ works positively for DM 
detection.}
\end{figure}

Now in Fig.~\ref{fig:crosssec} we show an advantage, from the point of view of constraining DM parameters, of considering 
IMBH minispike enhancements. We show contributions to the EGB at an energy of 1 GeV, against the DM annihilation cross-section. 
As before, we use $z_{re}=10.9$ and $\beta=0.3$, and assume for now $m_\chi=100\,\mathrm{GeV}$. The advantage of minispikes 
is observed in the gradient of the plotted line. Although one would na\"\i vely expect gamma-ray fluxes to scale as 
$\sigma\mathit{v}/m_\chi^2$, the presence of a minispike alters this dependence. This is due to the fact that the dominant 
term in the minispike enhancement [the integral in Eq.~(\ref{delta2})] is given by the expression $\sim \rho_{lim}^2r_{lim}^3$, 
which brings the gamma-ray flux scaling as $(\sigma\mathit{v})^{2/7}m_\chi^{9/7}$, for minispikes growing out of a $\gamma=1$ 
profile. Physically speaking, a smaller cross-section works to maintain a denser minispike, which compensates for the decrease 
in flux due to the smaller cross-section. One can say this weak dependence on DM parameters make minispikes particularly suited 
for DM detection.

To determine the potential of the weak dependence on DM parameters, we consider gamma-ray observations by GLAST, which is expected 
to have more than an order better point source sensitivity than EGRET. With its launch, many gamma-ray sources that could not have 
been resolved until now will be detected, and taking these into account, a smaller EGB flux is expected. The most widely considered 
candidate for the dominant EGB contributor is unresolved blazars, i.e.~a beamed population of active galactic nuclei (see 
\cite{NarumotoTotani} and references therein), and the fraction of these blazars that can be removed with GLAST depends on their 
luminosity function. The latest calculation predicts that the resolvable fraction is around 20\% \cite{NarumotoTotani}. If the EGB 
flux is entirely due to blazars, then the EGB will be reduced by the same fraction, and hence the sensitivity to $\sigma \mathit{v}$
will be improved by a factor of 2. We should, however, keep in mind that \cite{NarumotoTotani} concluded that blazars cannot fully 
explain the EGB flux, but only 25--50\%. Therefore, the remaining 50--75\% may be due to other astrophysical objects of either 
known or unknown origin, which may or may not be resolved by GLAST. This indicates that a significant fraction may still be resolved 
with GLAST, depending on the property of this additional contributor. In an optimistic case, where a fraction of 
$0.75+0.25\times 0.2 = 0.8$ can be resolved with GLAST, the cross-section sensitivity will be around 
$3 \times 10^{-29}\,\mathrm{cm^{-2}\,s^{-1}}$, depending on values of $z_f$ and $N_{bh}$ (see error bars in Fig.~\ref{fig:crosssec}). 
The sensitivity will also depend on $M_{bh}$; if halve $M_{bh}$, the sensitivity decreases to $\sim 10^{-28}\,\mathrm{cm^{-2}\,s^{-1}}$. 
These sensitivities are so small that no other experiments can compete in the next decades.

\subsection{Line Gamma-Rays}

\begin{figure}
\includegraphics[scale=0.4]{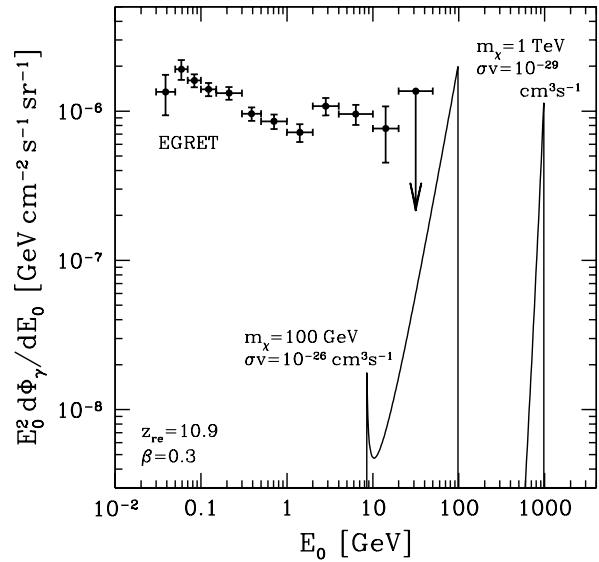}
\caption{\label{fig:linegamma}Spectral signature in the EGB due to DM annihilation into monochromatic photons in scenario B IMBH 
minispikes. The $2\gamma$ branch is considered here, with a fixed branching ratio of $10^{-3}$. Two DM parameters sets are shown, 
$m_\chi=100\,\mathrm{GeV}$ and $\sigma\mathit{v}=10^{-26}\,\mathrm{cm^3\,s^{-1}}$ (this is not excluded by the continuum component), 
and $m_\chi=1\,\mathrm{TeV}$ and $\sigma\mathit{v}=10^{-29}\,\mathrm{cm^3\,s^{-1}}$. We used $z_{re}=10.9$ and $\beta=0.3$.}
\end{figure}

In the previous subsection we discussed constraining DM parameters using contributions to the EGB from DM annihilation into 
continuous photons, and highlighted its potential using the future GLAST mission. However, the continuum component lacks a 
distinguishing signature to separate it from other sources, such as unresolved blazars. Although the spectrum of other sources 
decreases rapidly at high energies, this is nonetheless a difficulty. Ideally, we would like to identify DM annihilation without 
these complications. The monochromatic component of DM annihilation provides a means of achieving this. As discussed earlier, 
the monochromatic emission provides a `smoking gun' signature of DM annihilation due to its energy at the DM mass. Here, we 
present the monochromatic spectrum from the $\chi\overline{\chi} \to 2 \gamma$ process with a fixed branching ratio of 
$b_{2\gamma}=10^{-3}$. Our result is shown in Fig.~\ref{fig:linegamma} for two sets of DM parameters, $m_\chi=100\,\mathrm{GeV}$ 
and $\sigma v=10^{-26}\,\mathrm{cm^3\,s^{-1}}$ (note this is not excluded by the continuum component), and $m_\chi=1\,\mathrm{TeV}$ 
and $\sigma v=10^{-29}\,\mathrm{cm^3\,s^{-1}}$. Again, we consider scenario B IMBH minispikes with $z_{re}=10.9$ and $\beta=0.3$.  

The spectral shape of Fig.~\ref{fig:linegamma} arises because of distortion due to cosmological redshift and absorption of gamma-rays 
during propagation. The larger peak at $E_0=m_\chi$ is characteristic of DM annihilation, and if detected provides very convincing 
evidence for DM. The smaller peak is characteristic of IMBH minispikes; it is due to the strong enhancement factor at $z_f$, as 
shown in Fig.~\ref{fig:enhancement}. If detected, this second peak identifies the presence of a high enhancement factor at high 
redshift, supporting the case of seed-BHs and minispikes.

In the past, line gamma-rays had not been intensively considered, because the EGB spectrum taken by EGRET did not reach up to the 
required $O(100)$ GeV energies. With GLAST however, the energy window extends up to 300 GeV (and with better energy resolution), 
and line gamma-rays could be a serious candidate of the first DM detection. Furthermore, as shown in Fig.~\ref{fig:linegamma}, 
the minispike around IMBHs provide such high EGB line fluxes that it might give better evidence than the continuum flux. The 
characteristic spectral feature, combined with good energy resolution of GLAST, work quite positively for the detection of line 
gamma-rays, even if gamma-rays from other astrophysical sources give considerable contribution at the same energy.

\section{\label{sec:conc}Discussion and Conclusions}

We have studied contributions to the EGB from DM annihilation in minispikes around IMBHs. Our results are plotted in 
Fig.~\ref{fig:flux} using DM parameters $m_\chi=100\,\mathrm{GeV}$ and $\sigma\mathit{v} = 3 \times 10^{-26}\,\mathrm{cm^3\,s^{-1}}$. 
We find that a consideration of minispikes increase the contribution by $1-3$ orders, so that in optimistic scenarios the predicted 
gamma-ray flux may reach current EGB values. The EGB can therefore potentially be used to constrain DM parameters and/or IMBH 
scenarios, particularly when better EGB observations are taken by GLAST. In our work, we considered two IMBH formation scenarios, 
scenario A being remnants of Population-III stars \cite{scenarioA}, and scenario B being formed in the centres of protogalactic 
disks \cite{scenarioB}. Scenario A is our conservative case (mass $10^2M_\odot$ with a mild $r^{-3/2}$ minispike) while scenario 
B is our optimistic (mass $10^5M_\odot$ with a $r^{-7/3}$ minispike). 

We also showed that DM annihilation into monochromatic gamma-rays may be a serious contender for indirect DM detection. Our result, 
for scenario B IMBHs and two sets of DM parameters, is shown in Fig.~\ref{fig:linegamma}. Note that these parameters were chosen 
so that it is not excluded by the continuum component. The higher energy peak, at an energy equivalent to the DM mass, is within 
GLAST's potential observation range, and can provide a potential `smoking-gun' signature in the EGB. 

Compared to SMBH spikes, IMBH minispikes have the disadvantage that it is smaller, and it grows out of less dense DM profiles. 
Also, the survival probability of spikes and minispikes are still uncertain. However, there are still advantages to using IMBH 
minispikes. First, it has been shown that observations of our galaxy's centre constrain the strength of spikes to such a degree 
that DM annihilation in spikes cannot significantly contribute to the EGB \cite{Ando}. Second, although survival probabilities 
are uncertain, we can select IMBHs that are likely to have maintained their minispikes. Our selection involves choosing IMBHs 
that \emph{have not experienced any mergers}, on the grounds that mergers strongly destroy DM enhancements. Additionally, such 
unmerged IMBHs are likely to reside in the outskirts of galactic halos, where it is not affected by dynamical process that are 
also known to deplete DM enhancements.

It must also be added that minispike formation requires some initial conditions to be met, including an adiabatic and symmetric 
BH formation, formation of a BH in the centre of its host halo, and very cold initial DM orbits near the halo centre. These 
conditions are generally supported by the collisionless nature of particle DM, and adiabaticity is satisfied when one compares 
the BH formation time-scale to the dynamical time-scale at some relevant distance \cite{BertoneZentnerSilk}. However, the collapse 
and accretion processes during BH formation are not well known, and are likely to be complex and far from symmetric. Also, we 
have neglected the effects of seed-BHs born with enough kick-recoil velocities to be expelled out of their host halos. Although 
a detailed study is beyond the scope of this paper, minispike formation for such seeds may be surpressed.


An advantage of DM detection using minispikes is the fact that the gamma-ray flux is weakly dependent on DM parameters. This is 
because the smaller the $\sigma\mathit{v}$, the longer the minispike remains sharp. This fact compensates for the lack in flux, 
bringing the flux proportional to $(\sigma\mathit{v})^{2/7}m_\chi^{9/7}$. As a consequence, GLAST may be able to probe down to 
$\sigma\mathit{v}\sim10^{-29}\,\mathrm{cm^3\,s^{-1}}$ if we optimistically assume scenario B IMBH minispikes, and that 80\% of 
the EGB is resolvable by GLAST. This is far smaller than any other experiment, and excludes the allowed $\sigma\mathit{v}$ region 
due to the standard thermal relic SUSY DM scenario. However, evidence concerning IMBH properties and EGB composition are required 
before setting such constraints. In this respect, it has recently been shown that the EGB power spectrum can be used to discriminate 
DM contributions from other sources \cite{AndoKomatsu}.

All our calculations can be applied to other numerous DM candidates by simply substituting the appropriate differential gamma-ray 
spectrum per annihilation Eq.~(\ref{annihilationspectrum}). A detailed search over particle DM candidates using EGB contributions 
should become possible with the launch of GLAST, and these should then be cross-correlated with constraints from other potential DM 
signals. Finally, we have used EGB contributions to reveal DM properties, but we stress that the fact that scenario B predicts 
gamma-ray fluxes that are two orders greater than scenario A may also shed light on IMBH and SMBH models.

\begin{acknowledgments}
We thank Gianfranco Bertone for discussions.
S.A.~was supported in part by Sherman Fairchild fund at Caltech and
by a Grant-in-Aid from the JSPS.
\end{acknowledgments}

\bibliography{HoriAndo}

\begin{thebibliography}{10}

\bibitem{JungmanKamionGriest}
G.~Jungman, M.~Kamionkowski and K.~Griest,
\newblock Phys. Rep. {\bf 267}, 195 (1996).

\bibitem{Bergstrom}
L.~Bergstrom,
\newblock Rep. Prog. Phys. {\bf 63}, 793 (2000).

\bibitem{BertoneHooperSilk}
G.~Bertone, D.~Hooper and J.~Silk,
\newblock Phys. Rep. {\bf 405}, 279 (2005).

\bibitem{GLAST}
http://glast.stanford.edu/.

\bibitem{Bengtsson}
H.~U. Bengtsson, P.~Salati and J.~Silk,
\newblock Nucl. Phys. {\bf B346}, 129 (1990).

\bibitem{Berezinsky}
V.~Berezinsky, A.~Bottino and G.~Mignola,
\newblock Phys. Lett. B {\bf 325}, 136 (1994).

\bibitem{BergstromUllio}
L.~Bergstrom and P.~Ullio,
\newblock Nucl. Phys. {\bf B504}, 27 (1997).

\bibitem{BergstromUllioBuckley}
L.~Bergstrom, P.~Ullio and J.~H. Buckley,
\newblock Astropart. Phys. {\bf 9}, 137 (1998).

\bibitem{BergstromEdsjoGunnarsson}
L.~Bergstrom, J.~Edsjo and C.~Gunnarsson,
\newblock Phys. Rev. D {\bf 63}, 083515 (2001).

\bibitem{Cesarini_etal}
A.~Cesarini, F.~Fucito, A.~Lionetto, A.~Morselli and P.~Ullio,
\newblock Astropart. Phys. {\bf 21}, 267 (2004).

\bibitem{Hooper_etal}
D.~Hooper, I.~de~la Calle~Perez, J.~Silk, F.~Ferrer and S.~Sarkar,
\newblock JCAP {\bf 0409}, 002 (2004).

\bibitem{Fornengo_etal}
N.~Fornengo, L.~Pieri and S.~Scopel,
\newblock Phys. Rev. D {\bf 70}, 103529 (2004).

\bibitem{Horns}
D.~Horns,
\newblock Phys. Lett. B {\bf 607}, 225 (2005).

\bibitem{GondoloSilk}
P.~Gondolo and J.~Silk,
\newblock Phys. Rev. Lett. {\bf 83}, 1719 (1999).

\bibitem{UllioZhaoKamion}
P.~Ullio, H.~S. Zhao and M.~Kamionkowski,
\newblock Phys. Rev. D {\bf 64}, 043504 (2001).

\bibitem{Merritt_etal}
D.~Merritt, M.~Milosavljevic, L.~Verde and R.~Jimenez,
\newblock Phys. Rev. Lett. {\bf 88}, 191301 (2002).

\bibitem{BertoneMerritt}
G.~Bertone and D.~Merritt,
\newblock Phys. Rev. {\bf D72}, 103502 (2005).

\bibitem{Merritt}
D.~Merritt,
\newblock Phys. Rev. Lett. {\bf 92}, 201304 (2004).

\bibitem{ZaharijasHooper}
G.~Zaharijas and D.~Hooper,
\newblock Phys. Rev. D {\bf 73}, 103501 (2006).

\bibitem{MillerColbert}
M.~C. Miller and E.~J.~M. Colbert,
\newblock Int. J. Mod. Phys. {\bf D13}, 1 (2004).

\bibitem{BertoneZentnerSilk}
G.~Bertone, A.~R. Zentner and J.~Silk,
\newblock Phys. Rev. D {\bf 72}, 103517 (2005).

\bibitem{Sreekumar_etal}
EGRET, P.~Sreekumar {\em et~al.},
\newblock Astrophys. J. {\bf 494}, 523 (1998).

\bibitem{StrongMoskalenkoReimer}
A.~W. Strong, I.~V. Moskalenko and O.~Reimer,
\newblock Astrophys. J. {\bf 613}, 956 (2004).

\bibitem{BergstromEdsjoUllio}
L.~Bergstrom, J.~Edsjo and P.~Ullio,
\newblock Phys. Rev. Lett. {\bf 87}, 251301 (2001).

\bibitem{Ullio}
P.~Ullio, L.~Bergstrom, J.~Edsjo and C.~Lacey,
\newblock Phys. Rev. D {\bf 66}, 123502 (2002).

\bibitem{TaylorSilk}
J.~E. Taylor and J.~Silk,
\newblock Mon. Not. R. Astron. Soc. {\bf 339}, 505 (2003).

\bibitem{ElsasserMannheim}
D.~Elsaesser and K.~Mannheim,
\newblock Astropart. Phys. {\bf 22}, 65 (2004).

\bibitem{ElsasserMannheim2}
D.~Elsaesser and K.~Mannheim,
\newblock Supersymmetric dark matter and the extragalactic gamma ray
  background, 2005.

\bibitem{Ando}
S.~Ando,
\newblock Phys. Rev. Lett. {\bf 94}, 171303 (2005).

\bibitem{OdaTotaniNagashima}
T.~Oda, T.~Totani and M.~Nagashima,
\newblock Astrophys. J. {\bf 633}, L65 (2005).

\bibitem{AndoKomatsu}
S.~Ando and E.~Komatsu,
\newblock Phys. Rev. D {\bf 73}, 023521 (2006).

\bibitem{scenarioA}
P.~Madau and M.~J. Rees,
\newblock Astrophys. J. {\bf 551}, L27 (2001).

\bibitem{scenarioB}
S.~M. Koushiappas, J.~S. Bullock and A.~Dekel,
\newblock Mon. Not. R. Astron. Soc. {\bf 354}, 292 (2004).

\bibitem{Makishima_etal}
K.~Makishima {\em et~al.},
\newblock Astrophys. J. {\bf 535}, 632 (2000).

\bibitem{Fabbiano}
G.~Fabbiano,
\newblock Ann. Rev. Astron. Astrophys. {\bf 27}, 87 (1989).

\bibitem{Begelman}
M.~C. Begelman,
\newblock Astrophys. J. {\bf 568}, L97 (2002).

\bibitem{King_etal}
A.~R. King, M.~B. Davies, M.~J. Ward, G.~Fabbiano and M.~Elvis,
\newblock Astrophys. J. {\bf 552}, L109 (2001).

\bibitem{Mushotzky}
R.~Mushotzky,
\newblock Prog. Theor. Phys. Suppl. {\bf 155}, 27 (2004).

\bibitem{MillerFabianMiller}
J.~M. Miller, A.~C. Fabian and M.~C. Miller,
\newblock Astrophys. J. {\bf 614}, L117 (2004).

\bibitem{Cropper_etal}
M.~Cropper {\em et~al.},
\newblock Mon. Not. R. Astron. Soc. {\bf 349}, 39 (2004).

\bibitem{StrohmayerMushotzky}
T.~E. Strohmayer and R.~F. Mushotzky,
\newblock Astrophys. J. {\bf 586}, L61 (2003).

\bibitem{Fan_etal}
SDSS, X.~Fan {\em et~al.},
\newblock Astron. J. {\bf 122}, 2833 (2001).

\bibitem{Barth_etal}
A.~J. Barth, P.~Martini, C.~H. Nelson and L.~C. Ho,
\newblock Astrophys. J. {\bf 594}, L95 (2003).

\bibitem{WillottMcLureJarvis}
C.~J. Willott, R.~J. McLure and M.~J. Jarvis,
\newblock Astrophys. J. {\bf 587}, L15 (2003).

\bibitem{HaimanLoeb}
Z.~Haiman and A.~Loeb,
\newblock Astrophys. J. {\bf 552}, 459 (2001).

\bibitem{Fan_etal2}
SDSS, X.-H. Fan {\em et~al.},
\newblock Astron. J. {\bf 131}, 1203 (2006).

\bibitem{IslamTaylorSilk}
R.~R. Islam, J.~E. Taylor and J.~Silk,
\newblock Mon. Not. R. Astron. Soc. {\bf 340}, 647 (2003).

\bibitem{KoushiappasZentner}
S.~M. Koushiappas and A.~R. Zentner,
\newblock Astrophys. J. {\bf 639}, 7 (2006).

\bibitem{VolonteriHaardtMadau}
M.~Volonteri, F.~Haardt and P.~Madau,
\newblock Astrophys. J. {\bf 582}, 559 (2003).

\bibitem{LISAobs}
T.~Matsubayashi, H.~Shinkai and T.~Ebisuzaki,
\newblock Astrophys. J. {\bf 614}, 864 (2004).

\bibitem{FryerWoosleyHeger}
C.~L. Fryer, S.~E. Woosley and A.~Heger,
\newblock Astrophys. J. {\bf 550}, 372 (2001).

\bibitem{BondArnettCarr}
J.~R. Bond, W.~D. Arnett and B.~J. Carr,
\newblock Astrophys. J. {\bf 280}, 825 (1984).

\bibitem{NakamuraUemura}
F.~Nakamura and M.~Umemura,
\newblock Astrophys. J. {\bf 548}, 19 (2001).

\bibitem{ZhaoSilk}
H.~S. Zhao and J.~Silk,
\newblock Phys. Rev. Lett. {\bf 95}, 011301 (2005).

\bibitem{hydrogencooling}
M.~Tegmark {\em et~al.},
\newblock Astrophys. J. {\bf 474}, 1 (1997).

\bibitem{Zentner_etal}
A.~R. Zentner, A.~A. Berlind, J.~S. Bullock, A.~V. Kravtsov and R.~H. Wechsler,
\newblock Astrophys. J. {\bf 624}, 505 (2005).

\bibitem{SalamonStecker}
M.~H. Salamon and F.~W. Stecker,
\newblock Astrophys. J. {\bf 493}, 547 (1998).

\bibitem{NFW}
J.~F. Navarro, C.~S. Frenk and S.~D.~M. White,
\newblock Astrophys. J. {\bf 462}, 563 (1996).

\bibitem{MooreProfile}
B.~Moore, T.~Quinn, F.~Governato, J.~Stadel and G.~Lake,
\newblock Mon. Not. Roy. Astron. Soc. {\bf 310}, 1147 (1999).

\bibitem{MerrittGrahamMoore}
A.~W. Graham, D.~Merritt, B.~Moore, J.~Diemand and B.~Terzic,
\newblock astro-ph/0509417.

\bibitem{DiemandMooreStadel}
J.~Diemand, B.~Moore and J.~Stadel,
\newblock Mon. Not. Roy. Astron. Soc. {\bf 353}, 624 (2004).

\bibitem{Navarro_etal}
J.~F. Navarro {\em et~al.},
\newblock Mon. Not. Roy. Astron. Soc. {\bf 349}, 1039 (2004).

\bibitem{FukushigeKawaiMakino}
T.~Fukushige, A.~Kawai and J.~Makino,
\newblock Astrophys. J. {\bf 606}, 625 (2004).

\bibitem{AbelBryanNorman}
T.~Abel, G.~L. Bryan and M.~L. Norman,
\newblock Science {\bf 295}, 93 (2002).

\bibitem{BrommCoppiLarson}
V.~Bromm, P.~S. Coppi and R.~B. Larson,
\newblock Astrophys. J. {\bf 564}, 23 (2002).

\bibitem{Peebles}
P.~J.~E. Peebles,
\newblock Gen. Rel. Grav. {\bf 3}, 63 (1972).

\bibitem{QuinlanHernquistSigurdsson}
G.~D. Quinlan, L.~Hernquist and S.~Sigurdsson,
\newblock Astrophys. J. {\bf 440}, 554 (1995).

\bibitem{Merritt_book}
D.~Merritt,
\newblock Single and binary black holes and their influence on nuclear
  structure,
\newblock 2003, [astro-ph/0301257].

\bibitem{PressSchechter}
W.~H. Press and P.~Schechter,
\newblock Astrophys. J. {\bf 187}, 425 (1974).

\bibitem{ShethTormen}
R.~K. Sheth, H.~J. Mo and G.~Tormen,
\newblock Mon. Not. R. Astron. Soc. {\bf 323}, 1 (2001).

\bibitem{Jenkins_etal}
Virgo Consortium, A.~Jenkins {\em et~al.},
\newblock Astrophys. J. {\bf 499}, 20 (1998).

\bibitem{Page_etal}
L.~Page {\em et~al.},
\newblock astro-ph/0603450.

\bibitem{NarumotoTotani}
T.~Narumoto and T.~Totani,
\newblock Astrophys. J. {\bf 643}, 81 (2006).

\end{thebibliography}
\bibliographystyle{h-physrev4}   

\end{document}